\documentstyle{mn}

\def\hone{H\,{\sc i}}

\def\spose#1{\hbox to 0pt{#1\hss}}
\def\simlt{\mathrel{\spose{\lower 3pt\hbox{$\mathchar"218$}}
     \raise 2.0pt\hbox{$\mathchar"13C$}}}
\def\simgt{\mathrel{\spose{\lower 3pt\hbox{$\mathchar"218$}}
     \raise 2.0pt\hbox{$\mathchar"13E$}}}

\def\<{\thinspace}
\def\today{\ifcase\month\or January\or February\or March\or April\or May\or
      June\or July\or August\or September\or October\or November\or December\fi
      \space\number\day, \number\year}

\def\kms{{\rm\,km\,s^{-1}}}

\input{epsf}
\begin{document}

\title[Spectroscopy of the optical Einstein ring 0047-2808]
{Spectroscopy of the optical Einstein ring 0047-2808
\thanks{Based on observations obtained at the European
Southern Observatory, La Silla, Chile, and the United Kingdom
Infrared Telescope, Hawaii.}}

\author[S. J. Warren {\rm et al.}]
{S. J. Warren,$^{1}$ A. Iovino$^{2}$, P. C. Hewett,$^{3}$
and P. A. Shaver$^{4}$\\
$^{1}$Blackett Laboratory, Imperial College of Science Technology and
Medicine, Prince Consort Rd, London SW7 2BZ \\
$^{2}$Osservatorio Astronomico di Brera, Via Brera 28, I--20121 Milano, Italy \\
$^{3}$Institute of Astronomy, Madingley Road,
Cambridge CB3 0HA \\
$^{4}$European Southern Observatory, Karl--Schwarzschild--Strasse 2, 
D--85748 Garching bei M\"{u}nchen, Germany}

\date{Received
      in original form	}

\maketitle

\begin{abstract}

We present optical and near--infrared spectroscopic observations of the
optical Einstein ring 0047--2808. We detect both [OIII] lines
$\lambda\lambda$4959, 5007 near $\sim 2.3\mu$, confirming the redshift
of the lensed source as $z=3.595$. The Ly$\alpha$ line is redshifted
relative to the [OIII] line by $140\pm20\kms$. Similar velocity shifts
have been seen in nearby starburst galaxies. The [OIII] line is very
narrow, $130\kms$ FWHM. If the ring is the image of the centre of a
galaxy the one-dimensional stellar velocity dispersion $\sigma=55\kms$
is considerably smaller than the value predicted by Baugh et al. (1998)
for the somewhat brighter Lyman--break galaxies. The Ly$\alpha$ line is
significantly broader than the [OIII] line, probably due to resonant
scattering. The stellar central velocity dispersion of the early--type
deflector galaxy at $z=0.485$ is $250\pm30\kms$. This value is in good
agreement both with the value predicted from the radius of the Einstein
ring (and a singular isothermal sphere model for the deflector), and
the value estimated from the $D_n-\sigma$ relation.

\end{abstract}
\begin{keywords}
gravitational lensing -- galaxies: formation
\end{keywords}

\section{Introduction}

Warren et al. (1996a, hereafter Paper I) reported observations of the
first candidate Einstein ring to be discovered at optical
wavelengths. Confirmation that the system is a gravitational lens, via
the detection of a second emission line which secured the source
redshift, was recorded as a note added in proof. Here we present the
confirmatory spectrum together with additional optical and near-ir
spectra of the source and deflector galaxies.  The lens system,
designated 0047-2808, is the first example of a normal galaxy lensing
another normal galaxy, albeit one at high--redshift and probably
representing an early phase in the evolution of normal galaxies. The
deflector is a massive early--type galaxy, redshift $z=0.485$, and the
source a star--forming object at redshift $z=3.595$. Resolved
observations of the extended surface brightness distribution around the
ring using the Hubble Space Telescope offer the prospect of providing
powerful constraints on models of the deflector and the source in a
fashion similar to that achieved with Einstein rings discovered at
radio wavelengths (Kochanek 1995).

The detection of 0047-2808 highlights a number of important advantages
of the survey technique (Warren et al. 1996b) employed in the system's
identification. The deflector population, luminous early--type galaxies
with redshifts $0.3 \le z \le 0.5$, is homogeneous and well--defined,
and the deflector redshift is readily obtained. The source population,
optically faint high--redshift star--forming systems that exhibit
emission lines, notably Ly$\alpha$ $\lambda 1216$, is of particular
interest in charting the history of star--formation associated with the
population of normal galaxies, and, as with the deflectors, the object
redshift is readily obtained. The acquisition of redshifts for
deflector and source means that the lensing geometry is fully defined
and contrasts with the difficulties experienced in establishing the
redshifts of source and deflector in lenses discovered at radio
wavelengths (see summary by Keeton and Kochanek 1996).

In Section~\ref{optobs} new optical spectroscopic observations are
described, and the procedure used to measure the velocity--dispersion
of the deflector galaxy detailed. Infrared spectroscopy of the
restframe wavelength region including the [OIII] $\lambda\lambda 4959,
5007$ emission lines is presented in
Section~\ref{irobs}. Section~\ref{disc} contains a discussion of the
implications of the results for the nature of the source and deflector
galaxies.
 
\section{OPTICAL SPECTROSCOPY} \label{optobs}

\subsection {Observations}

Optical spectra of 0047-2808 were obtained at the ESO NTT 
equipped with the EMMI instrument in spectroscopic mode, on 1995
October 1 UT. Atmospheric conditions were not ideal with the seeing
averaging $1\farcs2$ and a strong wind throughout the night.  A
$600\,$l/mm grating blazed at $6000$\AA \ was employed in the EMMI red
arm, with a Tektronix $2048 \times 2048$ CCD as the detector. The CCD
was binned by a factor of two in the dispersion direction, giving a
pixel size of $0\farcs27 \times 1.3$\AA, and a wavelength coverage of
$5400-6700$\AA. The slit was oriented at PA$=45^{\circ}$ to include the
bright part of the lensed ring.

Four $2400\,$s exposures of the 0047-2808 system were made through a
$1\farcs5$ slit. The resulting spectra have a resolution of $\sim
3.25$\AA \ and include the Ly$\alpha$ $\lambda$1216 emission line of
the source galaxy at $5589$\AA, and the $4300$\AA \ G--band
absorption feature in the deflecting galaxy at $\sim 6390$\AA.

The EMMI blue arm was used with a slit width of $0\farcs75$ to obtain
spectra centered on the G--band absorption feature of three stars to
act as velocity--dispersion templates. The stars, HD8779, HD203638,  and
HD223311, are of spectral type K0III, K0IV, and K4III respectively. The
resulting spectra have a resolution of $\sim 1.13$\AA \ and a
wavelength coverage that exceeded that of the equivalent rest--frame
region accessible in the observations of the lensing galaxy.

\subsection {Spectroscopic reductions}

Data reduction employed standard routines available in {\tt
IRAF}~\footnote{{\tt IRAF} is distributed by the National Optical
Astronomy Observatories, which are operated by the Association of
Universities for Research in Astronomy, Inc. under contract with the
National Science Foundation} for bias, dark, and flat field correction,
and the removal of cosmic rays.  Variance-weighted one dimensional
spectra were extracted using an aperture width of 16 pixels,
$4\farcs3$. The spectra were rebinned to a linear vacuum wavelength
scale using $1.3$\AA \ and $0.45$\AA \ pixels for the red and blue arms
respectively. The rms residuals for individual calibration lines about
the adopted wavelength solution were $0.13$\AA \ and $0.02$\AA \
respectively.

The galaxy spectra were calibrated onto a relative flux scale using spectra
taken of a flux standard. The four galaxy spectra were then coadded using
variance weighting to produce a combined spectrum, shown in Figure
1. The zero--point of the flux scale was determined by equating the flux
in the spectrum integrated over a broad $V$ filter band--pass with
the $V$ total magnitude of the galaxy measured in the direct CCD image of
Paper I.  The signal--to--noise ratio over the wavelength range
$6100-6700$\AA \ in the final coadded galaxy spectrum was ${\mathrm
\sim 7\, pix^{-1}}$. The signal--to--noise ratios over the wavelength
range $4100-4500$\AA \ in the three stellar template spectra were $\sim
400$ (HD8779), $\sim 100$ (HD203638), and $\sim 250$ (HD223311).

\begin{figure*}
\vspace{10.5cm}
\includegraphics{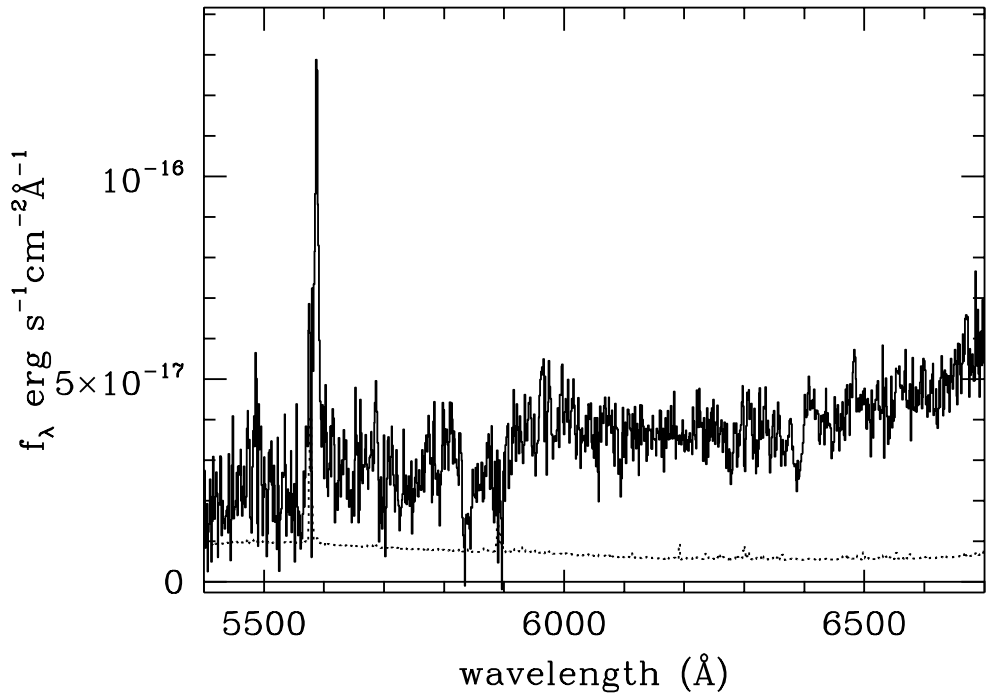}
\caption[]{Combined optical spectrum showing the G band (near $6400
{\rm \AA}$) and H$+$K (near $5900 {\rm \AA}$) absorption lines in
the deflector galaxy, and the Ly$\alpha$ emission line (at $5589 {\rm
\AA}$) from the source galaxy. The dotted line shows the 1 $\sigma$
error for each pixel. The pixel size is 1.31 \AA. The spectrum has been
scaled to the correct total magnitude.}
\end{figure*}

\subsection{Source emission line properties}

In order to measure the central wavelength and full width half maximum
(FWHM) of the Ly$\alpha$ $\lambda$1216 emission line we first
subtracted a redshifted and scaled template spectrum of an elliptical
galaxy. A low--order polynomial was then fitted to regions on either
side of the line in order to remove any residuals and a Gaussian
profile was then fitted, weighted by the errors as a function of
wavelength. The central wavelength was then corrected to the
heliocentric value.  The results are listed in Table 3.  The values are
in good agreement with those listed in Paper I, measured from a lower
resolution and lower signal--to--noise ratio spectrum.

\begin{figure*}
\vspace{10.5cm}
\includegraphics{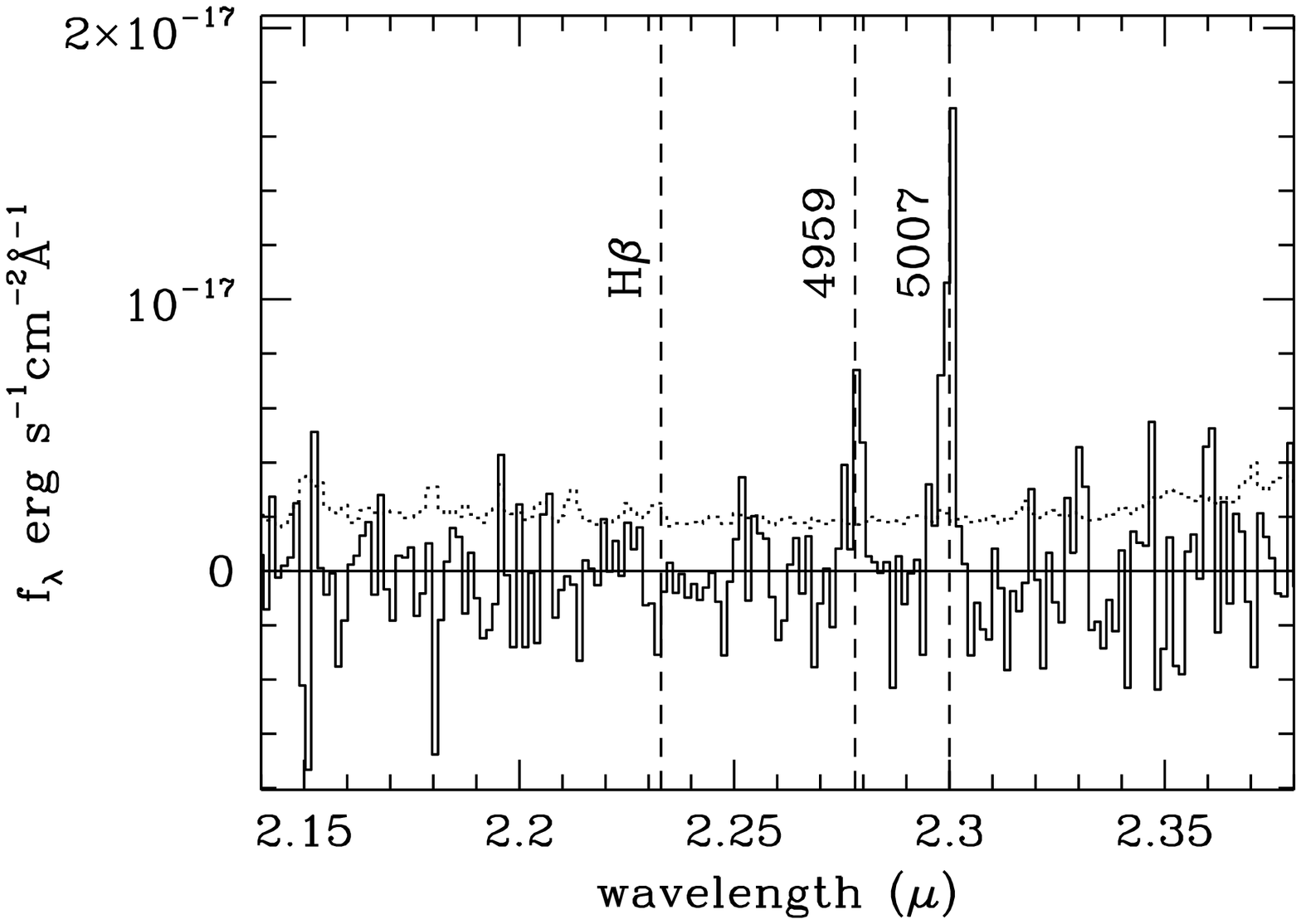}
\caption[]{Combined near-infrared spectrum showing the detected [OIII]
$\lambda\lambda$ 4959 and 5007 lines. This spectrum is an optimal
combination of the 4 spectra A, B, C, D, after subtraction of the
continuum. The spectra were rebinned to the largest pixel size of the 4
spectra ${\mathrm 0.0014\mu pix^{-1}}$ wide. The dotted line shows the
$1\sigma$ error for each pixel. The position of H$\beta$ is marked but
the summed flux at this wavelength is negative.}
\end{figure*}

\subsection{Deflector velocity dispersion}

The Fourier cross--correlation technique of Tonry and Davis (1979) was
employed to determine the velocity dispersion of the lensing galaxy,
restricting the wavelength coverage to the restframe region
$4100-4500$\AA. Measurements using the {\tt FXCOR} task within the
{\tt IRAF} package {\tt RV} were calibrated using the procedure
described by Falco et al. (1997) to which the reader is referred for
further details. In outline, a cross--correlation between the galaxy
and stellar--template spectra is performed after first smoothing the
template spectra to the same resolution as the galaxy spectrum
(FWHM$\sim 160 \kms$ at $\sim 4300$\AA \ restframe in our case).  A
Gaussian is then fit to the central portion of the resulting
cross--correlation peak to give a measure of the FWHM.  The galaxy
velocity dispersion is estimated via a calibration curve relating input
velocity dispersion to measured FWHM. The calibration curve is created
by convolving each stellar template with a series of Gaussians of
different width, and cross correlating against the original
template. For the continuum subtraction, apodizing, and filtering we
followed closely the recommendations of Tonry and Davis (1979).

Table~1 lists, for each template spectrum, the measured FWHM of the
cross--correlation peak, the inferred $\sigma_{v}$ from the associated
calibration curve, and the R--value, a measure of the signal--to--noise
ratio of the correlation peak (Tonry and Davis 1979).

The template that most closely matches the galaxy spectrum in the
strength of the absorption features is HD8779, spectral type K0III,
while HD223311, with the rather late spectral type of K4III, gives the
lowest value for $\sigma_{v}$. Assigning equal weight to the value
obtained using each template and taking the standard error among the
three measurements leads to an estimate of $249 \pm 10 \kms$.  However,
the signal--to--noise ratio of the galaxy spectrum is low and varying
the parameters that control the filtering of the Fourier components
produces differences of $\simlt 20\kms$. Varying the range over which
the Gaussian fit to the cross--correlation is performed by $\pm10\%$
also produces differences of $\simlt 20\kms$. Therefore we adopt $250
\pm 30 \kms$ as the estimate of the galaxy velocity dispersion.
  
\begin{table}
 \centering
  \caption{Velocity dispersion estimates}
  \begin{tabular}{@{}lccc}
   \hline \\[-12pt]
   Template & FWHM & $\sigma_{v}$ & R--value \\
            & km s $^{-1}$ & km s $^{-1}$ &  \\
   \hline \\[-12pt]
   HD8779   &  578  & 250  & 5.8 \\
   HD203638 &  600  & 260  & 5.8  \\
   HD223311 &  582  & 237  & 5.9  \\
   \hline
\end{tabular}
\end{table}

\section{NEAR-INFRARED SPECTROSCOPY} \label{irobs}

\subsection{Observations}

Spectra in the $K$ window covering the wavelengths of redshifted
H$\beta$ $\lambda$4861 and [OIII] $\lambda\lambda$4959, 5007 in the
lensed source were obtained in 1995, 1996, and 1997 using the CGS4
instrument on the United Kingdom Infrared Telescope. The journal of
observations appears in Table 2. As with the optical observations the
slit was aligned in a manner so as to cover the region of emission
where the ring is bright. For the 1995 and 1996 observations the slit
was centered on the galaxy oriented at the position angles (PAs) listed in
Table 2 i.e. pointing at the region where the ring is bright.
For the 1997 observations the slit was instead centered on the region
where the ring is bright and aligned tangentially to the ring.

\begin{table*}
 \centering
  \caption{Journal of near-infrared spectroscopy}
  \begin{tabular}{@{}lcccccccc}
   \hline \\[-12pt]
\multicolumn{1}{c}{UT date} & spectrum & slit width & grating & order 
& sub-step & resolving power & position angle & integration time \\
     & & arcsec    & l mm$^{-1}$  &  &  &  &  &   sec   \\
1995 Oct  1     & A & 2.4 &  75  & 1 & 2 & $\:450$ & 45$\:\:$ & 3520    \\
1996 Jul 31     & B & 2.4 & 150  & 1 & 1 & $\:900$ & 45$\:\:$ & 1920    \\
1996 Aug 1, 2   & C & 1.2 & 150  & 1 & 2 &    1640 & 45, 0    & 2520    \\
1997 Dec 6      & D & 1.2 & 150  & 2 & 1 &    3560 & $-$45    & 3840 \\
  
   \hline
\end{tabular}
\end{table*}

The observing procedure involved nodding the galaxy between two
positions on the slit. If necessary, as indicated in Table 2, at each
nod position the spectrum was sub-stepped by 0.5 pixel in the
wavelength direction to recover the full sampling of the undersampled
spectra by interleaving exposures. 

\subsection{Spectroscopic reductions}

The data reduction mostly followed the standard CGS4 methodology for
combining all the frames, with the exception of the use of a
sigma--clipping algorithm in summing the sets of co-added integrations
at each nod position (in order to improve the rejection of cosmic
rays), and an improved polynomial fit for second--order sky
subtraction. For the 1996 observations the failure of the CGS4 slit
rotation mechanism resulted in sky lines that were tilted away from the
perpendicular to the dispersion direction and it was necessary to apply
a distortion correction to straighten the frames.

A variance frame was created by measuring the variance in the counts in
each column, excluding the pixels containing the object spectrum, and
assuming this single value was applicable to each pixel within the
column. This estimate agreed well with the variance computed assuming
Poisson statistics. One-dimensional spectra were then extracted from
the data and variance frames, using apertures 1.2 arcsec wide for
spectra A, B, C, and 1.8 arcsec wide for spectrum D. The spectra from the
two nod positions were combined and the resulting spectrum rebinned
onto a linear vacuum wavelength scale. Finally the spectra were flux
calibrated using observations of standard stars.

\begin{figure}
\vspace{13.5cm}
\includegraphics{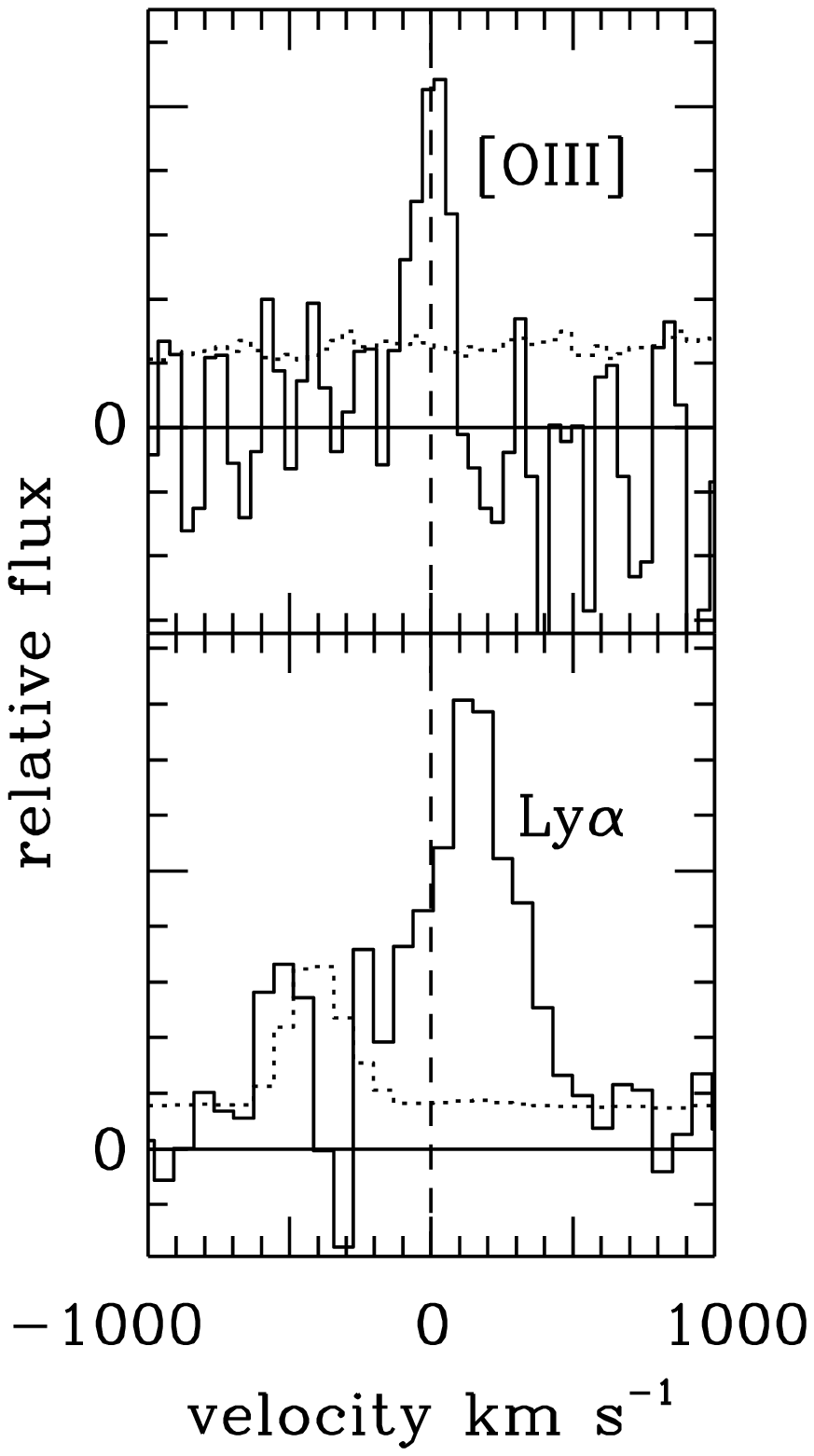}
\caption[]{Comparison of the redshift and width of the [OIII] 5007
(spectrum D) line and the Ly$\alpha$ line. The solid lines show the
continuum--subtracted spectra, and the dotted lines the $1\sigma$ error
spectra. Notice that the Ly$\alpha$ line is redshifted relative to the
[OIII] line and is also broader.}
\end{figure}

\begin{table*}
\begin{minipage}{140mm}
  \caption{Properties of the Ly$\alpha$, H$\beta$, and [OIII] 5007 lines}
  \begin{tabular}{@{}lccc}
   \hline \\[-12pt]
                            & Ly$\alpha$   & H$\beta$ &[OIII] 5007 \\
                            &              &   &            \\ \hline \hline
$^a$Central wavelength $\lambda$&${\mathrm 5589.1\pm0.2\AA}$& &${\mathrm
  2.3008 \pm0.0001\mu}$ \\
$^a$Redshift                    &$3.5974\pm0.0002$& &$3.5953\pm0.0002$ \\
$^b$Intrinsic FWHM km s$^{-1}$        &$390^{+50}_{-75}$& &$130^{+40}_{-60}$ \\
$^c$Line flux erg s$^{-1}$ cm$^{-2}$ &$(6.0\pm0.6)\times
  10^{-16}$& $<2.0\times  10^{-16}\: (3\sigma)$&$(6.0\pm0.8)\times  10^{-16}$ \\ \hline
\end{tabular}
\\
$^a$ vacuum heliocentric value \\
$^b$ FWHM errors are $2\sigma$, but $1\sigma$ for other parameters \\
$^c$ Ly$\alpha$ flux from narrow-band image (Paper I) \\
\end{minipage}
\end{table*}

\subsection{Source emission line properties}

The four K-band spectra are useful in a variety of ways. Because of the
wide slit used, the spectra A and B will provide reasonably accurate
measures of line fluxes (provided the surface brightness profile around
the ring in the light of [OIII] is similar to that in Ly$\alpha$).
Therefore spectra C and D, taken with a narrower slit, were scaled to
the correct total flux using the measured flux in the [OIII] 5007
line. In each spectrum a low--order polynomial fit was used to subtract
the continuum, and the four spectra were then averaged with weightings
proportional to the inverse variance. The final combined spectrum is
plotted in Fig. 2.

Table 3 lists the measured flux in the [OIII] 5007 line. Also provided
is the $3\sigma$ upper limit to the flux in the H$\beta$ line. The
summed flux over 5 pixels at the position of H$\beta$ is negative.  The
flux limit quoted is simply three times the $1\sigma$ error on the
summed flux at that wavelength i.e. we ignored that the measured flux
is negative. The central wavelength of the [OIII] 5007 line was
measured from spectra C and D, and corrected to the heliocentric value.
Spectra A and B are not useful for this measurement because the large
width of the slit means that the surface brightness of the source
galaxy may be non-uniform across the slit so that the centroid of the
flux may not coincide with the centre of the slit. Finally the width of
the [OIII] line was measured from spectrum D. The line is unresolved in
the other three spectra. The line width was measured by simultaneously
fitting Gaussians, of line--flux ratio 3:1, to the redshifted 5007 and
4959 lines. The value quoted in Table 3 is the intrinsic width
i.e. after quadrature subtraction of the instrument profile. A plot of
the velocity profile of the [OIII] 5007 line in spectrum D is provided
in Fig. 3. In the following section we compare the redshift and line
width of the [OIII] line measured from spectrum D with the
corresponding values for Ly$\alpha$. This assumes that we have sampled
the same part of the ring for both spectra. This is close to the truth,
since both spectra were centered on the SW quadrant of the ring. For the
Ly$\alpha$ line a section of the ring 1.5 arcsec long (the slit width)
is represented in the spectrum. For the [OIII] line, with the slit in a
perpendicular orientation, a section 1.8 arcsec long (the extraction
aperture) was sampled.

\section{Discussion} \label{disc}

\subsection{The source galaxy}

\subsubsection{Redshift of the source galaxy}

The primary result of the near--ir spectroscopy of the source galaxy is
the detection of two emission lines near $2.3\mu$ (Fig. 2). The lines are
redshifted ${\rm [OIII]}$ $\lambda\lambda$ 4959 and 5007 matching with
Ly$\alpha$ 1216 in the optical at $5589$\AA. This establishes the
redshift of the source as $z=3.595$ and confirms this system as a
gravitational lens.

In Fig 3 the velocity profiles of the [OIII] 5007 and Ly$\alpha$ lines
are compared. The velocity zero point for both plots is the redshift
measured from the [OIII] 5007 line. Two differences between the
lines are apparent. Firstly the Ly$\alpha$ line is substantially
broader than the [OIII] line. This is presumably due to resonant
scattering of the Ly$\alpha$ line and is considered further
below. Secondly it is noticeable that the Ly$\alpha$ line is redshifted
relative to the [OIII] line. The velocity difference is $140\pm20$ km
s$^{-1}$. This probably has two causes. Firstly a redward shift of the
centroid of the Ly$\alpha$ line has also been seen in those nearby
starburst galaxies which display Ly$\alpha$ emission (e.g. Lequeux et al
1995) and this is related to the kinematics of the gas in the starburst
galaxy and to the escape path of the resonantly--scattered Ly$\alpha$
photons. This mechanism may therefore be partly responsible for the velocity
shift seen in Fig. 3. In addition absorption of the blue wing of the
Ly$\alpha$ emission line by intervening cosmologically--distributed
Ly$\alpha$--forest clouds may contribute to the redward shift of
the line centroid. If so the intrinsic width of the Ly$\alpha$ line is
even larger than the value quoted in Table 3.

\subsubsection{Velocity dispersion of the source galaxy}

The second important result of the near--ir spectroscopy is the
measurement of the width of the [OIII] $\lambda\lambda$ 4959, 5007
lines. The intrinsic width of the [OIII] line (Table 3) is 130 km
s$^{-1}$ FWHM. This corresponds to a one-dimensional velocity
dispersion of only $\sigma=55$ km s$^{-1}$. Such a low value could be
explained if, for example, we are imaging a region of star formation in
a disk, away from the nucleus. On the other hand the high--redshift
galaxies discovered by Steidel et al (1996) are very compact so it is
quite likely that the ring is the image of the centre of a galaxy. In this
case the galaxy mass is very small compared to the mass of typical
galaxies today. This result accords in a general way with the
hierarchical picture of galaxy formation in which today's galaxies were
in several pieces at $z>3$.  However the measured velocity dispersion
then appears low in comparison with the results of the calculations of
Baugh et al (1998) who used a semi--analytic approach to predict the
velocity dispersion of the Lyman--break galaxies discovered by Steidel
et al (1996). Although the magnitude range they considered (${\mathrm
{\cal R}_{AB}<25})$ is somewhat brighter than the unlensed magnitude of
the source galaxy, ${\mathrm V\sim27}$ (Paper I), the predicted
velocity dispersions $>200$ km s$^{-1}$ are several times larger than
the value measured for the ring.

Clearly it is of interest to detect other lensed systems similar to
0047-2808 in order to measure the distribution of line widths. The
advantage of lensed systems is the flux amplification ($\sim 20$ for
$0047-2808$) enabling higher S/N measurements and the possibility of
observing fainter sources. These results will complement the work of
Pettini et al (1998) who have begun a programme of K--band measurements
of the line widths of the brightest Lyman--break galaxies.

\subsubsection{Dust in the source galaxy}

The width of the Ly$\alpha$ emission line $390^{+50}_{-75}$ km s$^{-1}$
FWHM ($2\sigma$ limits) is significantly greater than the width of the
[OIII] line $130^{+40}_{-60}$ km s$^{-1}$ FWHM. This is probably due to
resonant scattering of the Ly$\alpha$ photons which diffuse in
frequency over the multiple scatterings during their escape through a
cloud of neutral hydrogen. 

Because the escape path length of the Ly$\alpha$ photons is much
greater than for the continuum photons, the line is extinguished
selectively relative to the continuum.  We now show how it is possible
to combine the measured width and flux of the Ly$\alpha$ line with the
flux of one of the Balmer lines to estimate the dust--to--gas ratio in
the gas cloud. Consider a source of Ly$\alpha$ photons escaping from
the mid-plane of a plane-parallel slab of neutral hydrogen. The line
width is a measure of the face-on column density of the cloud, because the
broadening of the emerging Ly$\alpha$ line is a function of the number
of times the escaping photons are scattered. The line width is
proportional to the cube root of the column density (e.g. Adams
1972). The extinction is also related to the number of times the
photons are scattered, and to the dust--to--gas ratio. Therefore the
extinction of the line is related to the line width $\Delta v$ and the
dust--to--gas ratio. Since the extinction can be measured by comparing
the strength of the Ly$\alpha$ line to one of the Balmer lines, the
measurement of the Ly$\alpha$ line width yields a measurement of the
dust--to--gas ratio.

In detail, Warren and M\o ller (1996) provided the following expression
for the fraction of escaping Ly$\alpha$ photons:
\begin{equation}
f_{\rm e}={\rm sech}\lbrack(\Delta
v/147)^2(\sigma/10)^{-1}(k\delta/0.2)^{1/2}0.143\rbrack 
\end{equation}
where $k\equiv10^{21}(\tau_{_{\rm B}}/{\rm N_{\hone}})$ cm$^{-2}$ is
the dimensionless dust-to-gas ratio, $\delta$ is the ratio of the
absorption optical depth in the continuum near Ly$\alpha$ to that in
the B band $\tau_{_{\rm B}}$ (Charlot and Fall 1991), and $\sigma$ is
the velocity dispersion of the neutral gas, taken here to be
$10\kms$. The Case B ratio of the Ly$\alpha$ and H$\beta$ line fluxes
is $\sim 25$, while the measured ratio is $>3$ (Table 3). Therefore the
Ly$\alpha$ line has been extinguished by a factor $1/f_{\rm e}$ no more
than 8. Inserting this limit into the above equation we obtain a limit
to the dust--to--gas ratio $k\delta<1.5$.

This limit is not particularly stringent. For example Fall, Pei, and
McMahon (1989) obtained a measurement of $k\delta\sim0.2$ in
high--redshift damped Ly$\alpha$ clouds, from the reddening of the
background quasars. Nevertheless the method is an independent one and
the limit could be made lower by obtaining a deeper spectrum at
H$\beta$. For a lower--redshift galaxy the stronger H$\alpha$ line
could be used. In a galaxy where both H$\alpha$ and H$\beta$ could be
measured the Balmer decrement would provide a consistency check since
the line ratio can be predicted knowing both the inferred column
density and the dust--to--gas ratio.

\subsection{Velocity dispersion of the deflector galaxy}

In Paper I we estimated the value of the stellar central velocity
dispersion $\sigma_c$ of the lensing galaxy by two methods based on the
measured galaxy surface--brightness profile. The Faber--Jackson and
$D_n-\sigma$ relations yielded values of $\sigma_c=280 (260)
\pm50\kms$, and $\sigma_c=300 (265) \pm35\kms$ respectively, for
$q_0=0.1$ (0.5). The measured value of $\sigma_c=250 \pm30\kms$,
reported above, is consistent with both estimates. In Paper I we also
computed the lens velocity dispersion using the angular size of the
ring, and the singular isothermal sphere as a model for the mass
profile of the galaxy. This provides an estimate of the velocity
dispersion of the dark matter $\sigma_{_{DM}}$. The lens angular radius
of 1.35 arcsec implies a value $\sigma_{_{DM}}=270 (265)\kms$. The
measured value of $\sigma_c$ is also consistent with this estimated
value of $\sigma_{_{DM}}$. This result bears on the question of the
expected value of the ratio of $\sigma_{_{DM}}/\sigma_c$.  Kochanek
(1993) argued that the ratio will be close to one, rather than the
value $\sqrt(3/2)$ adopted by Turner, Ostriker, and Gott (1984). This
difference is important in making predictions of the expected number of
lenses, and their angular separations, using measured values of
$\sigma_c$ for nearby galaxies. As compared with the lower value,
adoption of the larger value leads to predictions of over twice as many
lenses, with separations $50\%$ larger (Kochanek 1993). The measured
ratio $\sigma_{_{DM}}/\sigma_c=1.08 (1.06) \pm0.13$, for $q_0=0.1$
(0.5), is consistent with both values.  It is feasible with an
8m--class telescope to measure $\sigma_c$ with sufficient accuracy to
provide unambiguous discrimination between the alternatives. Indeed it
would be possible to measure the velocity dispersion profile, which,
combined with a deep high--resolution image of the ring, could be used
to test detailed models of the stellar dynamics and mass distribution
in the deflector galaxy.

\subsection*{Acknowledgments}
We are particular grateful to J. Davies, T. Geballe, S. Leggett,
and S. Ramsay Howat who undertook the various UKIRT CGS4 observations
on our behalf as part of the UKIRT service programme. The authors
acknowledge the data analysis facilities provided by the Starlink
Project which is run by CCLRC on behalf of PPARC.

\bsp 

\end{document}